# Continually Working Multiwire Porous Detectors of Radiation*

**M.P.Lorikyan**

Yerevan Physics Institute, Brothers Alikhanian Str. 2, Yerevan, 375036 ARMENIA

E-mail: lorikyan@moon.yerphi.am

lorikyan@star.yerphi.am

Performance of the known porous multiwire radiation detectors is stable in the mode of pulsed voltage supply. The results of the study of multiwire porous detector operated at constant voltage supply are presented in this article. It is shown that the detector's performance is stable, and its spatial resolution $\Delta x$ β ± 60 μm.

* Work supported by the International Science and Technology Center

The discovery of the phenomenon of controllable drift and multiplication of electrons in porous dielectrics under the influence of external electric field [1-5] allowed us to create a new class of radiation detectors – porous detectors (PD). Porous emission detectors had been designed first, in which the primary-particle-induced secondary electrons after their multiplication within the same emitter escaped into vacuum and were detected. The secondary emission factor of porous emitter was 250 in case of passage of minimum ionizing particles [6], and several tousands in case of α-particle, when its energy loss was ~2-3 MeV [7, 8-11]. For more details see Lorikyan's review [12].

Porous detectors are representing a new radiation detection technology. The rise time and pulse width in such detectors is less than 1 ns and 2 ns, and the time and spatial resolution is 60 ps and 250 μm, respectively [13, 14]. These detectors are operated in a vacuum of β $10^{-2}$ Torr, have little quantity of matter ($5\times10^{-4}$ g/cm$^2$), and are highly efficient at detecting charged particles and soft X-rays. Such properties allow PD to be used in accelerator experiments, time-of-flight measurements, for detection of soft X-rays in space, and X-ray imaging in medicine.

The PDs work as follows: the ionization electrons (δ-electrons) produced in the walls of pores by the incident ionizing radiation are accelerated in the pores by the applied electric field. These electrons produce secondary emission electrons in the walls of other pores, which are in their turn accelerated in the pores, emitting new secondary electrons. This process occurs with all generations of secondary electrons and an avalanche multiplication of electrons is taking place in the porous medium with a high gain factor.

Although there is currently no qualitative theory of electrons drift and multiplication in porous media, one can conjecture that high σ is due to the fact, that the electron affinity of the pore material drops down to negative values, as a result of which the electrons are easily escaping from bottom of conduction band into vacuum. Also, electrons are escaping from regions deeper behind the surface of pore walls. So, the share of electrons produced by initial particles and secondary electrons taking part in the process of electron drift and multiplication becomes larger.

In earlier studies of multiwire porous detectors (MPD) [12-14, 15-18, 19-21] we observed that the radiation detection efficiency was rapidly decreasing with time. To ensure stable performance of detectors, we had to operate them in the mode of pulsed voltage supply [22] – narrow (2-5 ms) working voltage pulses alternating with wide (18-15 ms) depolarizing voltage pulses. Analysis of the foregoing results shows that polarization effects are arising due to the impurities that are present in the porous material [23].

The present study was conducted for MPD with CsI of 99,98 % purity. MPD was assembled under as clean technological conditions as possible.

The schematic view of MPD is shown in Fig.1.

Anode wires 1 with diameter 25 μm are stretched across the surface of fiberglass laminate frame 2. Porous CsI layer 6 is first deposited on aluminum cathode 4 of thickness 65 μm, then the frame with anode wires on it and frame 5 with the cathode fastened on it are strained against each other. Frame 3 is for setting the detector’s gap.

The MPD being used had a gap of 0,5 mm and a sensitive area of 22,0×22,0 $mm^2$. The anode wires were spaced at 0,25 mm. The thickness of deposited CsI layer was 0,75mm, i.e. the anode wires were buried in the porous layer.

Porous CsI was prepared by means of thermal deposition in an Ar atmosphere [40]. The CsI layer had a density of $34 \times 10^{-2}$ $g/cm^3$. The measurements were taken under (7-9)$\times 10^{-3}$ Torr vacuum. The X-rays incident through the cathode had 5,9 KeV energy and 63 $s^{-1}$ intensity, while α-particles getting in from the side of anode wires had 5 MeV energy and 48 $s^{-1}$ intensity. In all cases, the errors are taken to be statistical and are not indicated. The pulse registration threshold was 35 mV.

For estimation of coordinate resolution, the MPD's anode wires were grouped into two groups. The first group included odd-numbered anode wires, and the second group included the even-numbered anode wires. Anode wires in each group were connected to the inputs of fast current aplifiers having a conversion ratio of 30 mV/μA. After amplification, the signals from the first group, *N(I),* and second group, *N(II),* were shaped and counted. Signals from shaped pulses were also counted. $N_{cc}$ indicate that the same particle was detected by both groups, i.e., by two neighboring anodes. It is obvious that the ratio of $N_{cc}$ to the total number of detected particles is determined by the coordinate resolution of MPD.

The studies have shown that when the voltage to MPD was applied in less than some time $T$ after deposition of the CsI layer, both anode groups were registering all particles simultaneously, which means that the same particle had been registered by two neighboring anode wires simultaneously, i.e. MPD's spatial resolution was worse than the spacing between the anode wires, and the registration efficiency $\eta$ was decreasing with time. But when the MPD was turned on in time $T$ after deposition of the porous layer, one could see a completely different picture: the MPD had a good spatial resolution and its performance was stable in time. The investigations have shown that for MPD having porous CsI with $\rho \approx 0,4\%$, $T$ is 7 hours and for $\rho \approx 0,7\%$ $T$ = 5 hours.

Fig.2 shows the dependence of the number of α-particles registered by the first group $N_\alpha(I)$ (crosses), second group $N_\alpha(II)$ (points), both groups $N_{cc}$ (triangles) and only one group $N_\alpha(I)$ - $N_{cc}/2$ (squares) of anode wires on voltage, which was observed in 5 hours after MPD was assembled. The curve with triangles represents the number of cases when the same α-particle is registered by two groups of anode wires $(N_{cc})$.

One can see, that all the curves are forming a plateau, and, within the errors, each group of anode wires has counted the same number of α-particles, and the number of coincidence pulses $N_{cc}$ is very small compared with the number of α-particles registered by both groups of anode wires (≈ 3%), i.e. the spatial resolution of MPD is less than the spacing between the anode wires, i.e. better than ± 125 μm. Deterioration of coordinate resolution due to the parallax effect was < 60 μm on the average. In this case, the total number of registered particles $N_\alpha = N(I) + N(II) - N_{cc}$.

Fig. 3 shows the time-stability of $N_\alpha(I)$ (crosses), $N_\alpha(II)$ (points), $N_{cc}$ (triangles), and the number of α-particles detected by only one group of anode wires $N_x(I)$ - $N_{cc}/2$ (squares) measured at U=1538 V, right after measuring the dependence presented in Fig.2. It is obvious, that the time-stability of MPD is high. Later on, the same MPD was operated 6 days, for 15 hours a day, being turned off for the remaining 9 hours of the day. Analogous data were obtained. The detection efficiency sometimes dropped by about 10%, which was easily recovered by adjustment of the voltage. Similar results were obtained with the anode wires spaced at 125 μm, i.e. the MPD's spatial resolution was better than ± 60 μm.

The results of 15-hour time-stability measurements carried out on the sixth day are presented in Fig.4. The symbols used are the same as in all Figures. In the beginning of the measurements $U$=1400 V, but in three hours of operation the voltage was set at 1438 V. This Figure shows that on the first day of measurements $\eta$ and $N_{cc}$ remained unchanged within the errors.

Upon finishing the measurements, the detector was disassembled. Examination of MPD revealed no changes in the CsI layer, which means, that the MPD could have been still operable.

Analogous measurements were conducted with X-rays.

Fig.5 shows the results of the measurement of the dependence of the number of X-quanta registered by the first group $N_x(I)$, (crosses), second $N_x(II)$, (points), both groups $N_{cc}$ (triangles) and by only one group $N_x(I)$ - $N_{cc}/2$ (squares) of anode wires on voltage. Measurements were conducted in 5 hours after deposition of CsI layer. The results of the same kind of measurements conducted on the third day are shown in Fig.6 (no measurements on the second day). No curves for $N_{cc}$ are presented for this case, because $N_{cc}$ is small, comprising 0,7% of $N_x = N_x(I)+N_x(II) - N_{cc}$. Note, that $N_{cc}$ is in a strong dependence on the quality of MPD assembly and CsI deposition. The results of time-stability measurements carried out for 14 days are presented in Fig.7, where the number of days is laid off along the X-axis, and the average number of $N_x$ per day is laid off along the Y-axis. On the second, fourth and ninth days the MPD was off. The values of $U$ on the given day are shown on the plot. As is seen in Fig.7, the X-ray registration efficiency decreased by about 30 % on the third day of measurements then remained constant at this level within the errors. The slight increase of η after the tenth day is more likely due to incorrect adjustment of voltage $U$. $N_{cc}$ did not practically change over the whole period of the measurements.

The author is thankful to Ajvazyan G., Ajvazyan R., Vardanyan H. and Asryan G. for assistance in preparation of this work.

## Figure captions

**Fig.1** The schematic view of MPD: *1* - anode wires, *2* - fiber glass laminate frame, *3*- dielectric frame, *4* - cathode, *5* - supporting frame, *6* - porous CsI layer.

**Fig.2** The dependence of $N_\alpha(I)$ (crosses), $N_\alpha(II)$ (points), $N_{cc}$ (triangles) and $N_\alpha(I)$ - $N_{cc}/2$ (squares) on voltage, observed in 5 hours after MPD was assembled.

**Fig.3** The time stability of $N_\alpha(I)$ (crosses), $N_\alpha(II)$ (points), $N_{cc}$ (triangles) and $N_\alpha(I)$ - $N_{cc}/2$ (squares) after measuring the data shown in Fig. 2.

**Fig.4** The time stability of $N_\alpha(I)$ (crosses), $N_\alpha(II)$ (points), $N_{cc}$ (triangles) and $N_\alpha(I)$ - $N_{cc}/2$ (squares) on the sixth day.

**Fig.5** The dependence of the number of X-quanta $N_x(I)$ (crosses), $N_x(II)$ (points), $N_{cc}$ (triangles) and $N_x(I)$ - $N_{cc}/2$ (squares) on voltage, measured in 5 hours after deposition of CsI layer.

**Fig.6** The dependence of the number of X-quanta $N_x(I)$ (crosses), $N_x(II)$ (points), $N_{cc}$ (triangles) and $N_x(I)$ - $N_{cc}/2$ (squares) on voltage, obtained on the third day.

**Fig.7** The time stability of $N_x(I)$ (crosses), $N_x(II)$ (points), $N_{cc}$ (triangles) and $N_x(I)$ - $N_{cc}/2$ (squares) during 14 days.

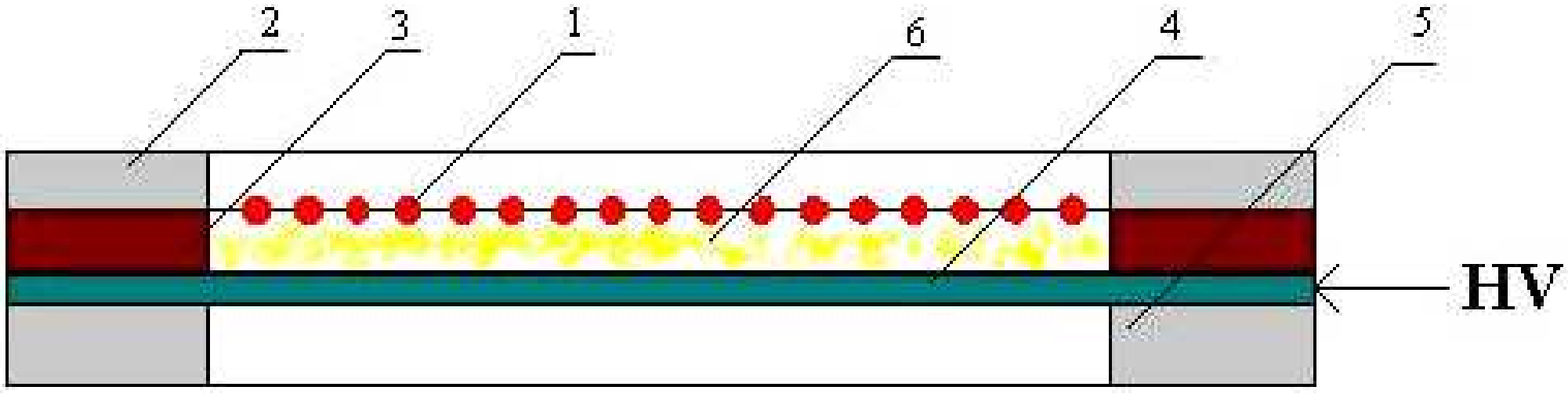

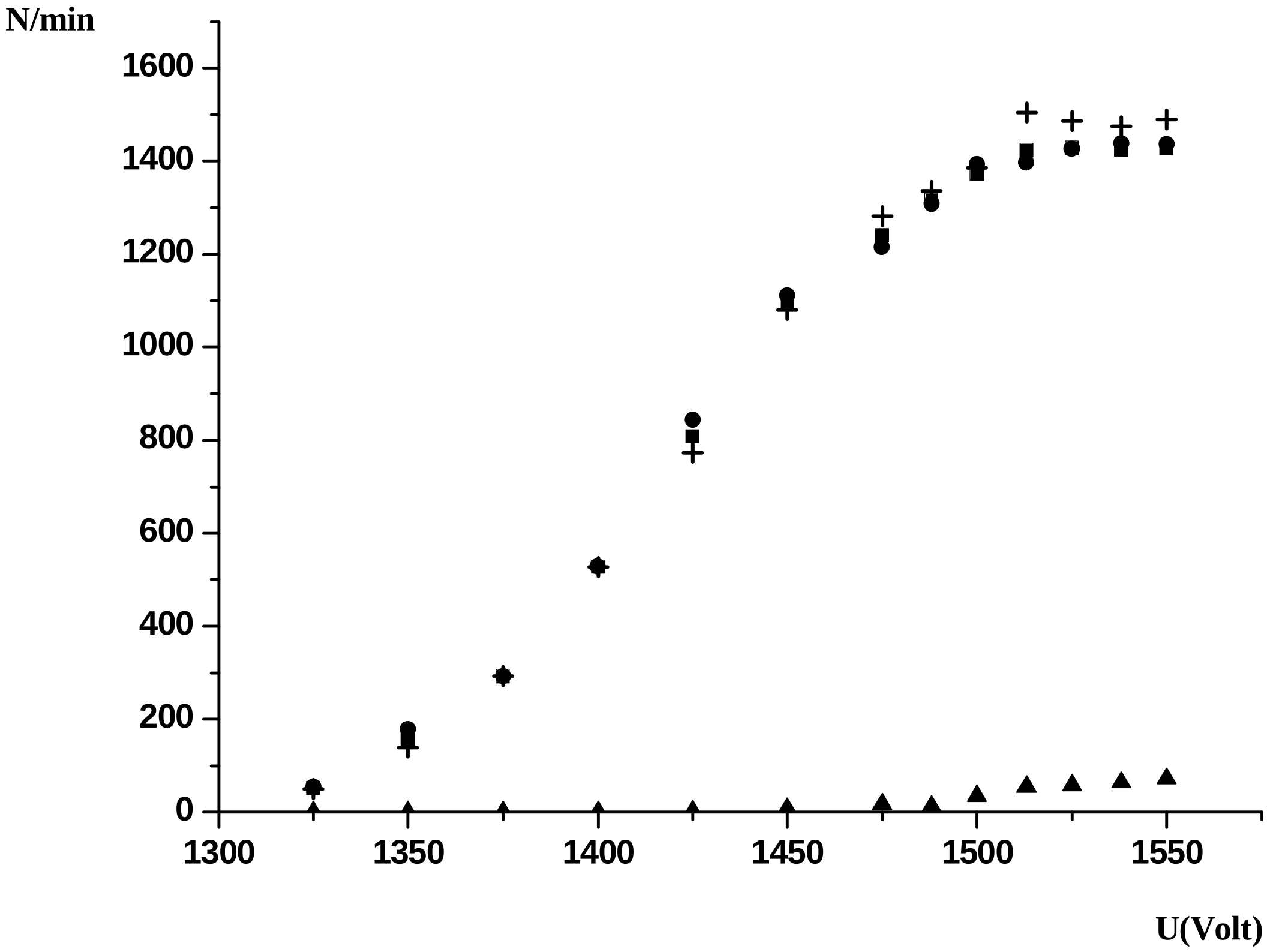

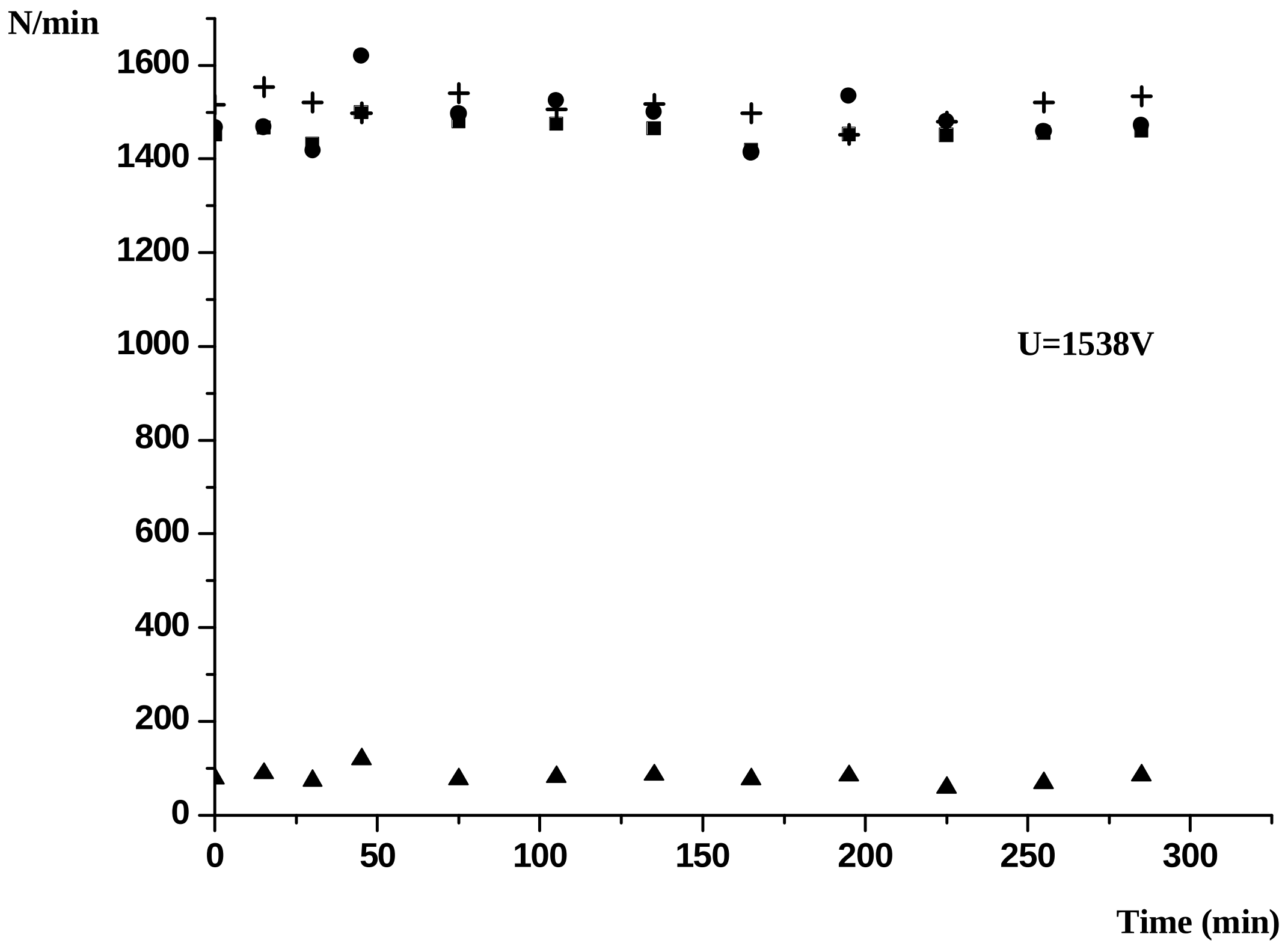

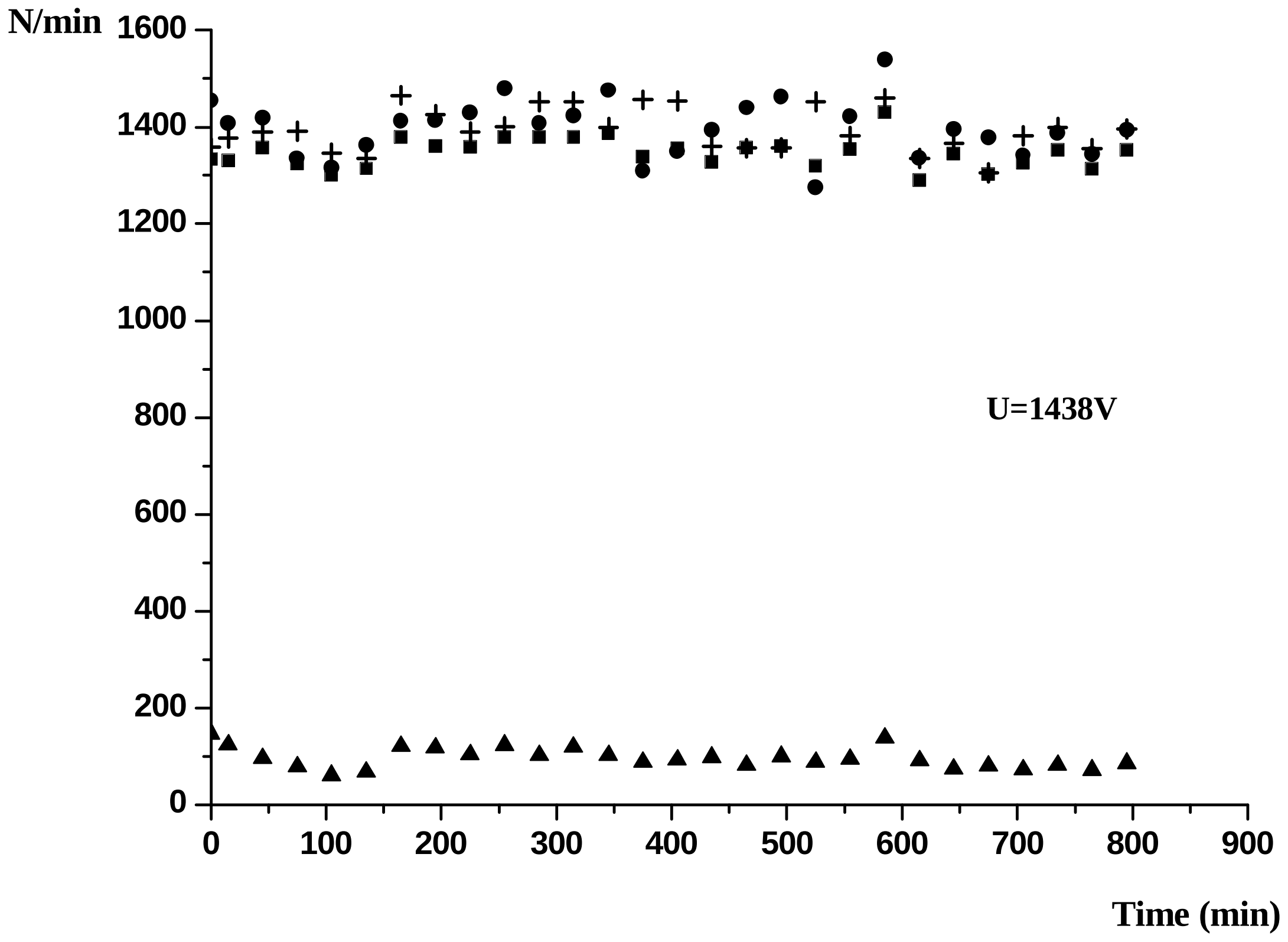

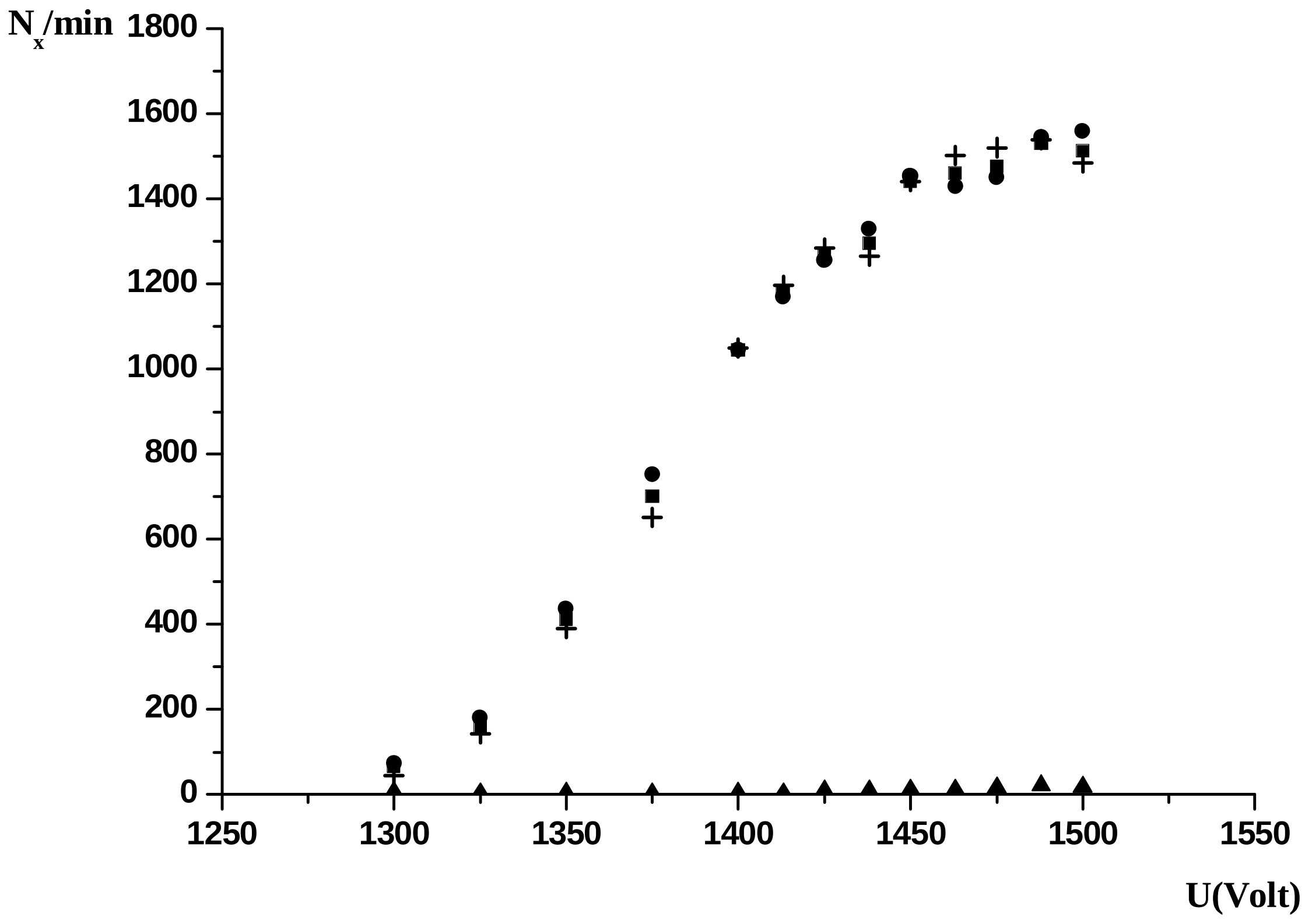

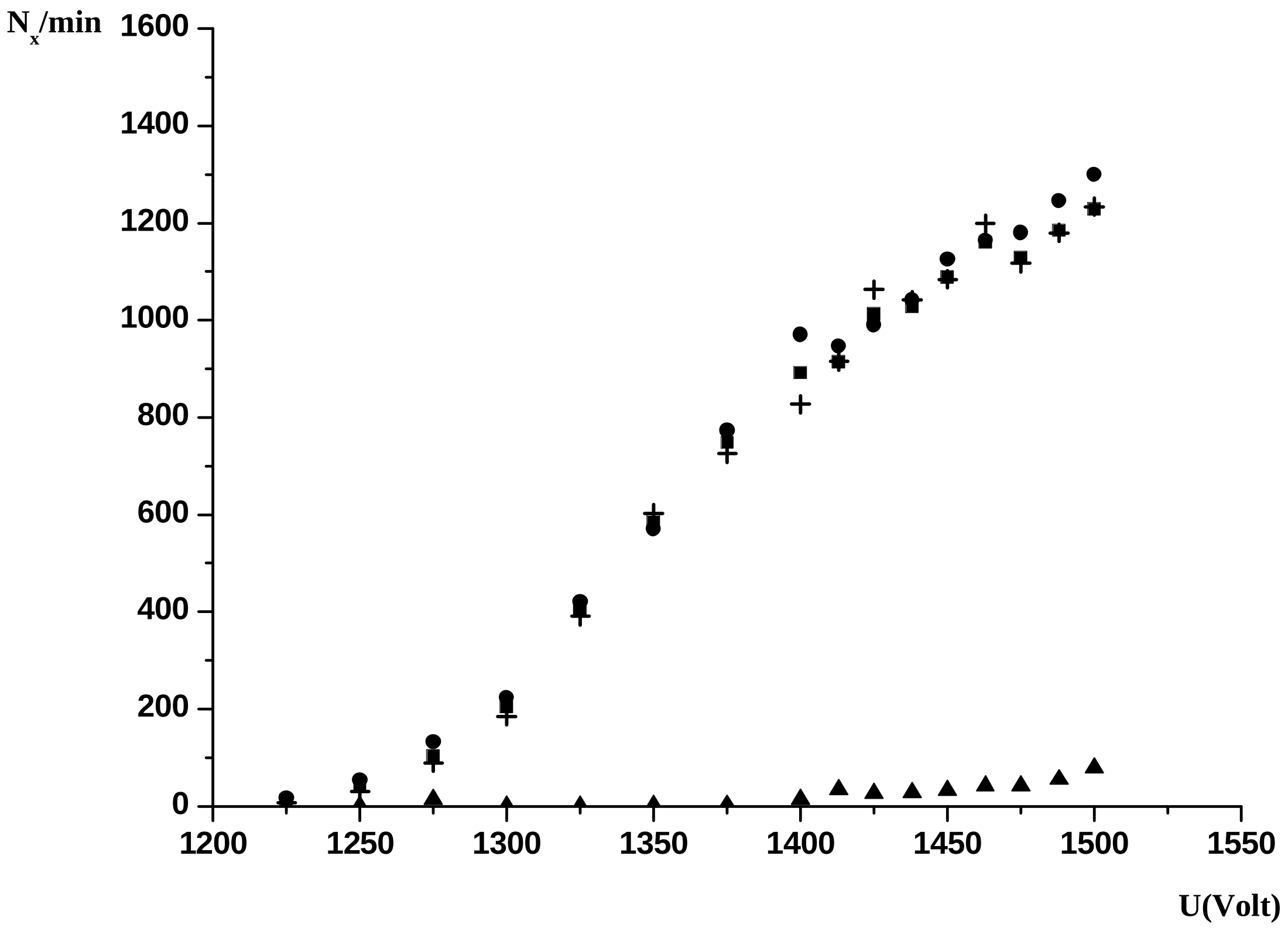

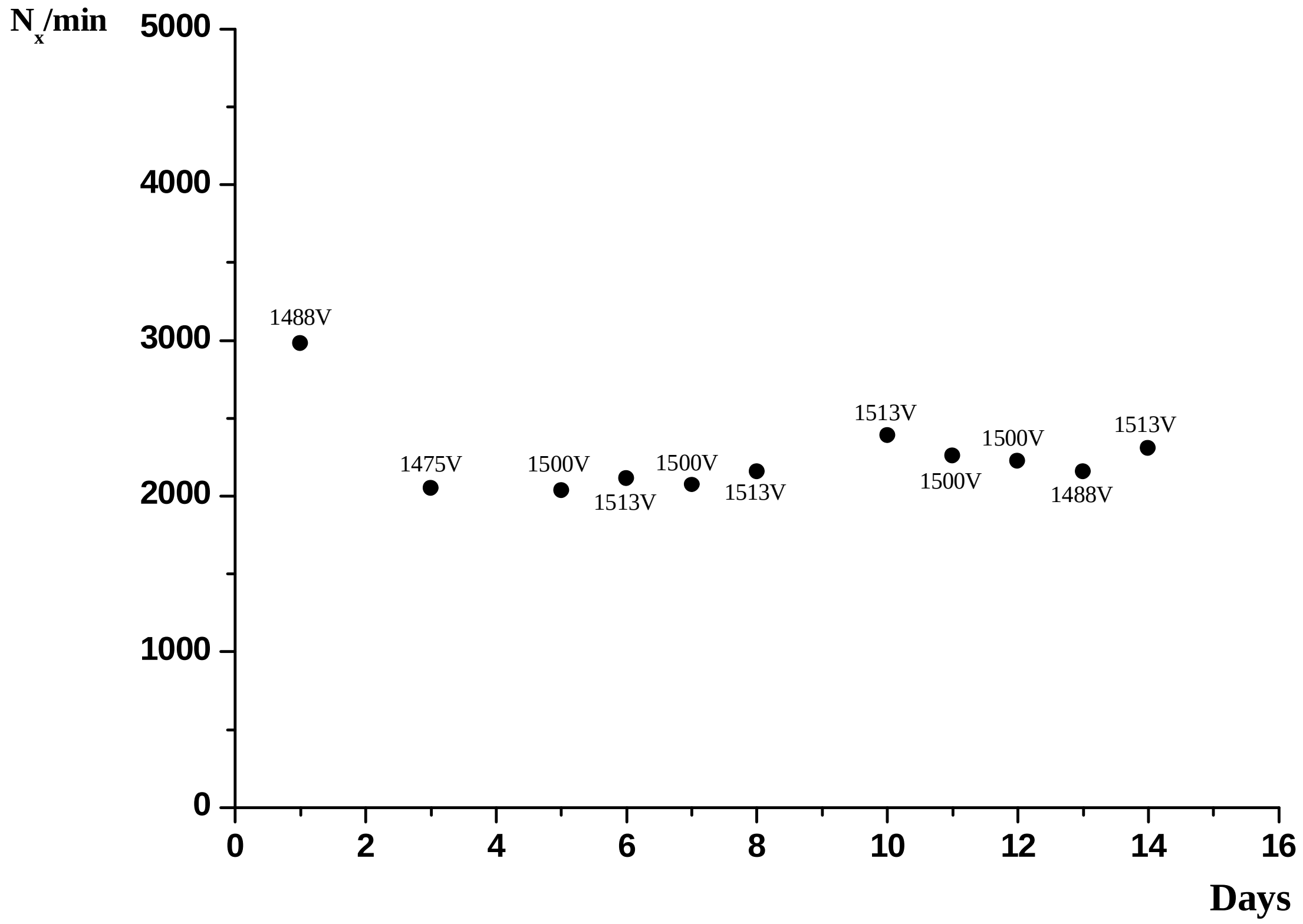